\documentclass{elsart}

\usepackage{epsf}
\usepackage{amssymb}

\begin{document}

\begin{frontmatter}

\title{Quantum mechanics without statistical postulates}

\author{H. Geiger\thanksref{emailharald}},
\author{G. Obermair\thanksref{emailgustav}} and
\author{Ch. Helm\thanksref{emailhelm}}

\address{ Institut  f\"ur Theoretische  Physik,  Universit\"at  Regensburg,
D-93040 Regensburg,  Germany }

\thanks[emailharald]{harald.geiger@physik.uni-regensburg.de}
\thanks[emailgustav]{gustav.obermair@physik.uni-regensburg.de}
\thanks[emailhelm]{christian.helm@physik.uni-regensburg.de}

\begin{abstract}
The Bohmian formulation of quantum mechanics is used in order to describe  
the measurement process in an intuitive way without a reduction postulate in
the framework of a deterministic single system theory. 
Thereby the motion of the hidden classical particle is
chaotic during almost all nontrivial measurement processes.
For the correct
reproduction of experimental results, it is further essential that the
distribution function $P(x)$ of the results of a position measurement
is identical with $\vert\Psi\vert^2$ of
the wavefunction $\Psi$ of the single system under consideration. It is shown that
this feature is not an additional assumption, but can be derived strictly
from the chaotic motion of a single system during a sequence of
measurements, providing a completely deterministic picture of the
statistical features of quantum mechanics. 
\end{abstract}


\end{frontmatter}

\section{Introduction}

Since the invention of quantum mechanics overwhelming experimental
support for this basic theory of nonrelativistic physics has been
accumulated.
In contrast to this great success, quantum reality created paradoxa or
counterintuitive behaviour from the very beginning, one of the most
prominent being
"Schr\"odinger's cat" and the problem of the  reality of a quantum state in 
the absence of  an observer. 
Most of these problems have their origin in the lack of a microscopic
description of the measurement process, where an ad hoc reduction of the
wavefunction $\Psi$ is assumed.

In order to avoid this assumption, a deterministic formulation of quantum
mechanics has been suggested bei David Bohm \cite{Bo52}. In this  theory 
the dynamics of a nonlocal hidden
variable is derived from the wavefunction $\Psi$. It is equivalent
 to the "standard" formulation of quantum
mechanics with respect to the prediction of experimental results, but allows
for an continuous and conceptually clear analysis of the measurement process without additional
assumptions.
Up to now it seemed that an additional statistical assumption
 concerning the distribution $\mathcal{P}(x,t)$ of the particles in an
 ensemble $\{ x_i \} $
has to be made, in order to reproduce the experimental results. 
This is one of the main reasons, why this intuitive classical interpretation of
quantum mechanics had been abandoned in the early days, as this assumption
is in contrast to a purely deterministic formulation \cite{Bo54,Ke53}. 

In the following it will be proven that this statistical assumption can be
derived strictly from the properties
 of this purely deterministic theory by considering the
chaotic dynamics of the Bohmian particle.
This shows that quantum mechanics can be understood completely on the basis
of a nonstatistical formulation.

The paper is organized as follows: After a short review of von Neumann's
description of the measurement process in standard quantum mechanics, this
approach is reconsidered from the point of view of the Bohmian
interpretation. 
By the investigation of the chaotic motion of the
particle it will be demonstrated that quantum equilibrium will be
established intrinsically during a sequence of measurements of a single
system.

\section{Standard quantum mechanics \label{standard}}

Within the standard interpretation of quantum mechanics the state of 
a system is completely
described by the wavefunction $\Psi(x,t)$ of the system  \cite{Ne32}.
The time evolution of the wavefunction is not only determined by the
unitary process according to the  Schr\"odingerequation 
\begin{equation} 
i\hbar \frac \partial {\partial t} \Psi(x,t) =
-\frac{\hbar^2}{2m} \frac{\partial^2}{\partial x^2} \Psi(x,t) +
V(x,t)\Psi(x,t) \label{Schr} \quad , 
\end{equation}
but also by the ad hoc reduction of the wavefunction in case of a
measurement.
In this process
the original
wavefunction $\Psi(x)=\sum_nc_n\Psi_n(x)$ which is a superposition of
eigenfunctions $\Psi_n(x)$ 
of the Observable $\hat A_x$, is replaced with probability $\vert
c_n\vert^2$ by  one eigenfunction $\Psi_{n_0}(x)$
with the eigenvalue $a_{n_0}$.
For the case of simplicity a discreet spectrum with nondegenerate
eigenvalues has been assumed.

In von Neumann's approach to the measurement process the measurement 
device  is described by a wavefunction $\Phi_0(y)$ with a
standard deviation $\sigma_{\Phi_0(y)}>0$ of $\vert\Phi_0(y)\vert^2$ 
and is considered as an integral part
of the total quantum system  \cite{Ne32}.
As there is no connection between the
measurement device and the system to be measured before the measurement, the
total initial wavefunction \begin{equation}
\Psi_0(x,y)=\Psi(x)\Phi_0(y) \label{psianfang}
\end{equation} is per definition a product state (Fig. 1(a)). 
During the measurement an interaction \begin{equation}
 \hat H_{WW}=\lambda \hat A_x\hat p_y
\end{equation}
($\hat p_y=-i\hbar \frac \partial {\partial y}$, $\lambda =const.$) 
between the detector (with coordinate $y$)  and the
system  ($x$) 
is assumed, which is strong and short compared with the interaction
$V(x,t)$  in the unperturbed Schr\"odinger equation \ref{Schr}. 
Therefore the dynamic of
the total system during the measurement process is governed by $\hat H_{WW}$ alone
and the time evolution of the total wavefunction is given by 
\begin{equation} 
\Psi(x,y,t)=\sum_nc_n\Psi_n(x)\Phi_0(y-\lambda a_n
t) \quad . \label{immer} 
\end{equation}
This shows that  the modulus  $\vert\Psi(x,y,t)\vert^2$ of the 
total wavefunction
separates into disjunct wavepackets along the detector coordinate $y$ during
the measurement process (Fig. 1(b)), provided that
$\lambda \Delta a \Delta t>\sigma_{\Phi_0(y)}$ ($\Delta t=$ 
duration of the measurement, 
$\Delta a=\min_n (a_n - a_{n-1})$). 
As each wavepacket corresponds to an eigenvalue $a_n$ of the 
observable $\hat A_x$, the choice of the measured quantity
influences the modification of the total wavefunction during the measurement
process.

\begin{figure}
\leavevmode
\begin{center}
\epsfxsize=\textwidth
\epsfbox{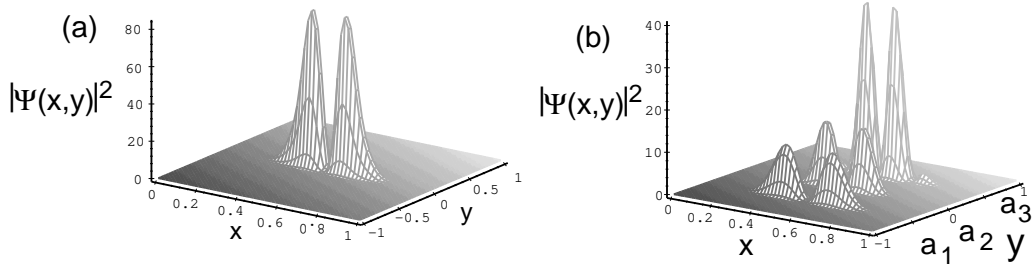}
\caption{ Schematic picture of a von Neumann measurement process: (a) At the beginning of the measurement 
the system is prepared in the product state $\Psi(x,y,0)=\Psi(x)\Phi_0(y)$ of
the system $\Psi(x)$ and the detector $\Phi_0(y)$ (cf. equ. \ref{psianfang}),
which is localized around $y=0$. 
(b) During the measurement process the wavepacket
$ \vert \Psi(x,y,t) \vert$ splits into 
different wavepackets along the $y$-direction (cf. equ. \ref{immer}), 
 corresponding to different eigenvalues $a_n$ of the
measured observable $\hat A_x$ of the system.  }
\end{center}
\end{figure}

Although this formalism is statistically in perfect agreement with
all known experiments, for the description of 
a single measurement also the reduction process has to be
understood.  
This process is intrinsically statistical in the sense that 
 the statistical distribution $|c_n|^2$ of the 
results $a_n$ can be calculated, 
but the outcome of a single measurement cannot be predicted in principle. 
This is the origin of the
 the randomness in the usual interpretation of quantum mechanics.
In addition to this, the fact that 
the behaviour of the system during the measurement
cannot be analysed further leads to the famous quantum
paradoxa.

\section{Bohmian quantum mechanics}

On the other hand, the Bohmian quantum mechanics is not a statistical, but a
single system theory with a principal lack of these problems.
Within the Bohmian mechanics, a state 
of a system is
completely determined not only by the wavefunction $\Psi$, but also  
by the position $x(t)=(x_1,\ldots,x_N)(t)$ of 
a hidden particle in the configuration
space of the whole system \cite{Bo52,Al94,Bo93,Cu96,Ge98,Ho93}.

The dynamics of the wavefunction $\Psi$ is determined from the
Schr\"odinger equation \ref{Schr} in the usual way, while
the dynamic of the particle is deduced from the wavefunction
$\Psi(x,t)$.
By introducing the modulus $R(x,t)$ and the phase $S(x,t)$ of the
wavefunction 
$ \Psi(x,t)=R(x,t) e^{\frac i \hbar S(x,t)}$
the Schr\"odinger equation (\ref{Schr}) can be rewritten as 
\begin{equation} -\frac{\partial }{\partial t}S(x,t) =\frac { (\frac \partial {\partial
x} S(x,t))^2 }{2m} - \frac{\hbar^2}{2m} \frac{\frac{\partial^2}{\partial
x^2} R(x,t)}{R(x,t)} + V(x,t) \; , \label{HJ}
\end{equation}

\begin{equation} \frac{\partial }{\partial t}R(x,t)^2 +\frac \partial {\partial
x}\Big(R(x,t)^2 \frac{\frac{\partial}{\partial
x}S(x,t)}{m}\Big)  = 0 \quad . \label{Ko}
\end{equation}

While
equation (\ref{Ko}) represents a continuity equation for
 $\vert\Psi(x,t)\vert^2$,
equation (\ref{HJ}) can be interpreted as a 
Hamilton Jacobi equation of a classical particle
with coordinate $x$ in the potential
$- \frac{\hbar^2}{2m} \frac{\frac{\partial^2}{\partial
x^2} R(x,t)}{R(x,t)} +V(x,t) $. 
As a consequence, the 
momentum $p(t)$ and the energy $E(t)$  of the particle are determined by the
phase $S(x,t)$ of the wavefunction $\Psi(x,t)$ at its position $x(t)$:
\begin{equation} p(t):=\frac\partial{\partial x} S(x,t)\vert_{x(t)}
\label{pH} \quad , \quad
E(t):=-\frac\partial{\partial t} S(x,t)\vert_{x(t)} \label{energie} \quad .
\end{equation}

\subsection{Measurement Process \label{measurment}}

The interpretation of the quantum mechanical reality in the framework 
of Bohmian quantum mechanics allows for an elegant and conceptually clear 
 description of a single measurement without the necessity of a reduction 
process \cite{Bo93,Ge98,Bo84}.
In the following a Bohmian extension of the von Neumann measurement process
explained above will be developed. Thereby not only the additional particle
has to be taken into account, but the whole concept of the measurement
process has to be reconsidered.

Without loss of generality it is assumed that in a first step (i)
there will be an interaction of the investigated system and
a microscopic
detector with one or a few degrees of freedom.
After that there will be a macroscopic measurement (ii) 
 of the state of the  microscopic detector after the interaction. 

i) The detector is described by a particle at the position 
$y(t)$ and a wavefunction  $\Phi_0(y)$ with a standard deviation 
$\sigma_{\Phi_0(y)}$
of  $\vert\Phi_0(y)\vert^2$.
At all times during the interaction an additional particle in the
configuration space of system $x$ and detector $y$ at the position $(x,y)(t)$ is
present in the Bohmian theory.
The evolution of the wavefunction $\Psi (x,y,t) $
during this process is the same as
in the standard von Neumann theory described above
(cf. chapter \ref{standard}).
From the knowledge of $(x,y)(0)$ and $\Psi(x,y,0)$ the values of $(x,y)(t)$
can be calculated with equation (\ref{pH}) at any time $t\in [0,\Delta t] $
during the measurement.
Due to the equation (\ref{pH}) the dynamics of the particle is exclusively 
determined by the local behaviour of the 
wavefunction $\Psi(x,t)$  at the position $x(t)$ of the particle.

The crucial point is that after the different wavepackets of
$\vert\Psi(x,y,t)\vert^2$ are separated in the configuration space of the
detector and the system at the time $\Delta t$ (cf. equ. \ref{immer}), 
the particle  $(x,y)(\Delta t)$ is influenced only locally by  one wavepacket 
$\vert\Psi_{n_0}\Phi_0(y-\lambda
a_{n_0}\Delta t)\vert^2$ belonging to the single eigenfunction
$\Psi_{n_0}(x)$ and the
eigenvalue $a_{n_0}$ respectively.

As a consequence of the separation of the wavefunction $\Psi$ into disjunct 
wavepackets, the result $a_{n_0}$ of the measurement of ${\hat A}_x$ is coded
in the position $(x,y)(\Delta t)$ of the particle in one of the subsets  
\begin{equation}
  \label{subsets}
  M_{n_0} := \{ (x,y) \mid  | \Psi_{n_0} (x) \Phi_0(y-\lambda
a_{n_0}\Delta t) |^2 \neq 0    \} 
\end{equation}
of the total phase space.

\begin{figure}
\leavevmode
\begin{center}
\epsfxsize=\textwidth
\epsfbox{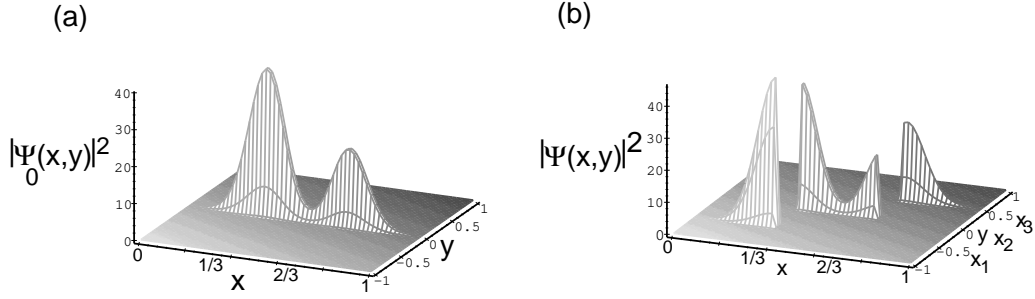}
\end{center}
\caption{Schematic picture of a position measurement with finite 
accuracy $\Delta x = \frac{1}{3}$ 
in the Bohmian formulation: 
(a) At the beginning of the position measurement the situation
is the same as in Fig. 1(a). 
(b) During the interaction of system and detector there is a separation of
wavepackets in $y$-direction. Each corresponds to one measurement
interval $[0,\frac 1 3[,[\frac 1 3,\frac 2 3[$ or $[\frac 2 3,1[$. The
deviation of the wavepacket is proportional to the position eigenvalue 
$x_i$. 
The position $(x,y) (\Delta t) \in M_{x_i}$ (cf. equ. \ref{subsets})
 of the particle at the end of the measurement
in one of the wavepackets determines the result $x_i$ of the experiment. 
}
\end{figure}

A special, but very important observable is the
position measurement, as any experiment involves at least the reading out of 
the position of some detector coordinate, e.g. of a pointer on a display
 \cite{Ne32,Sq90}.
The accuracy $\Delta x > 0$ of a 
position measurement is thereby in principle limited by the 
(unavoidable) standard deviation $\sigma_{\Phi_0 (y)}>0$
 of any real measurement apparatus.
During the interaction wavepackets, which correspond to different intervalls
$[x,x+\Delta x]$, are separated by different distances in the $y$-direction
of the detector (Fig. 2).

An important point is that the $x$-coordinate of the particle $(x,y)(t)$
does not change during the measurement, as the dominating interaction ${\hat
H}_{WW}$ only affects the wavefunction in the direction $y$.
As a consequence, the position $x(t)$, where the particle is detected during the
measurement, coincides with the initial position 
\begin{equation}
x(0) =x(t)=x(\Delta t) \label{ort} \quad \quad \quad 
\forall t  \in [0, \Delta t ] 
\quad .
\end{equation}

This is a nontrivial feature of the position measurement, which is not true for
arbitrary observables.


ii) For a complete measurement the information about the state of the
microscopic detector with a few degrees of freedom  has to be read out by a
macroscopic device. This is
described by a particle $z(t)=(z_1,\ldots, z_N)(t)$ and a wavefunction 
$\chi_0(z)$ with a standard deviation
$\sigma_{\chi_0(z)}$, $N$ being of the order of $10^{23}$.

The macroscopic limit of large $N$ has two important implications:
Firstly, a macroscopic experiment can be 
realized (without loss of generality) 
by a position measurement of the position $y$ via an interaction 
${\hat H}^{\prime}_{WW} = \lambda^{\prime} {\hat y} {\hat p}_z$. 
Due to the special
property (\ref{ort}) of the position measurement the result $a_{n_0}$ 
of the measurement (i)
contained in the coordinate $y$ (i.e. $(x,y) \in M_{n_0}$)
will not be affected by the manipulation (ii) along the $z$-coordinate 
described in the following.

Secondly on the macroscopic level the information about a measurement result
can be stored for a sufficiently long time, 
while in step (i) the wavepackets separated along $y$ can overlap again in
the course of the future dynamics. 
 If the overlap 
 $ \int dz_n \Psi_1 (z_n) \Psi_z (z_n)  \sim \epsilon < 1$
of the
wavepackets $\Psi_n$ in one dimension $z_i$ is small, the interference 
$\int d^N  z \Psi_1 (z) \Psi_z (z) $
of the complete wavefunctions in $N$ dimensions will be suppressed 
$ \sim \epsilon^N  \ll 1$. 
If the interference is missing in at least one dimension, the total overlap
is even exactly zero. 
This is a necessary condition for the storage of the information of the 
measurement (i) in a detector by the position of the particle $z(t)$,
 as in this case the particle trajectory is unable to leave the support 
  of the wavepacket corresponding to the eigenvalue 
$a_{n_0}$ (at least in $z$-direction) anymore (cf. equ. \ref{subsets}). 

Thus the Bohmian quantum mechanics describes the complete measurement in a
very elegant and clear way without the need of a reduction process.

\subsection{Deterministic Chaos\label{chaos}}

Chaotic phenomena within the Bohmian quantum mechanics have been
studied repeatedly
\cite{Sch95,Du92c,De96}. 
As it is possible to construct a trajectory
$(x(t),p(t))$ of the particle in the phase space of position and
momentum coordinates, deterministic chaos can be defined with the well known
Lyapunovexponent in the same way as in classical mechanics   \cite{Sch95}.
In the Bohmian quantum mechanics also the measurement is a deterministic
process, whose chaotic properties can be studied \cite{Ge98,Du92c,De96}.
During the first part (i) of the measurement process presented above the
dynamics of the system $x(t)$ and the detector $y(t)$ are given by the
system
\begin{equation}
\dot x(t)=\frac 1 {m_x}\frac \partial {\partial x}
S(x,y,t)\vert_{(x,y)(t)} \quad , \quad
\dot y(t)=\frac 1 {m_y}\frac \partial {\partial y}
S(x,y,t)\vert_{(x,y)(t)} \quad .
\end{equation}
of differential equations.
During a nontrivial measurement, which is connected with a modification of
$\Psi(x,y,t)$ according to equ. \ref{immer}, the phase $S(x,y,t)$ is time dependent.
 As the dimension of both the system and the detector is at least one,
 the Poincar\'e-Bendixson-Theorem 
\cite{Sch95}, which excludes chaos in autonomous systems of dimension
 $n\leq 2$,
is not applicable. Therefore in the general case chaotic dynamics of 
the hidden variable during the measurement process can be expected.

This behaviour becomes more explicit if 
 a sequence of measurements of a single system is considered, where
the information about the experimental
results is stored in a macroscopic device and the system is
repeatedly prepared in the initial states an infinite number of times.

For simplicity
a sequence of  position measurements with only two
intervals $[0,1[$ and $[1,2[$, 
for a wavefunction $\Psi(x)$ with a constant value of
$\vert\Psi(x)\vert^2$ in each interval is considered (cf. Fig. 3(a)). 
Because of the interaction of
the system with the detector the two wavepackets corresponding to the two
intervals separate from each other in the direction of the detector
coordinate $y$ (Fig. 3(b)) due to equation (\ref{immer}). 

Note that in the framework of Bohmian quantum mechanics the result $x_i$ 
of a
measurement is coded in the position of the particle $(x,y)  \in M_{x_i}$
in one of the separated
wavepackets corresponding to the interval $[x_i,x_i+\Delta x]$.
It is pointed out that this information will not be affected by the
following measurements, if it is preserved by the process (ii) in the 
$z$-coordinate of a macrocopic device. 
As the overlap of the wavepackets vanishes in at
least one dimension and the state $(x,y)$ of the system cannot leave this
sector of phase space anymore, the other part of configuration space is
irrelevant for all future dynamics and can therefore be neglected (Fig. 3(c)).
Thereby the measurement process breaks the ergodicity of the trajectory 
$(x,y)(t)$.
For a new preparation in the initial state (Fig. 3(a)) this remaining 
wavepacket has to flow into the form $|\Psi (x,y,t_0)|$ at $t=t_0$ (Fig. 3(d)).
This means that the phase space accessible for particle trajectories 
$(x,y) (t)$ will be enlarged along the $x$ coordinate
from Fig. 3(c) to 3(d).

\begin{figure}
\leavevmode
\begin{center}
\epsfxsize=\textwidth
\epsfbox{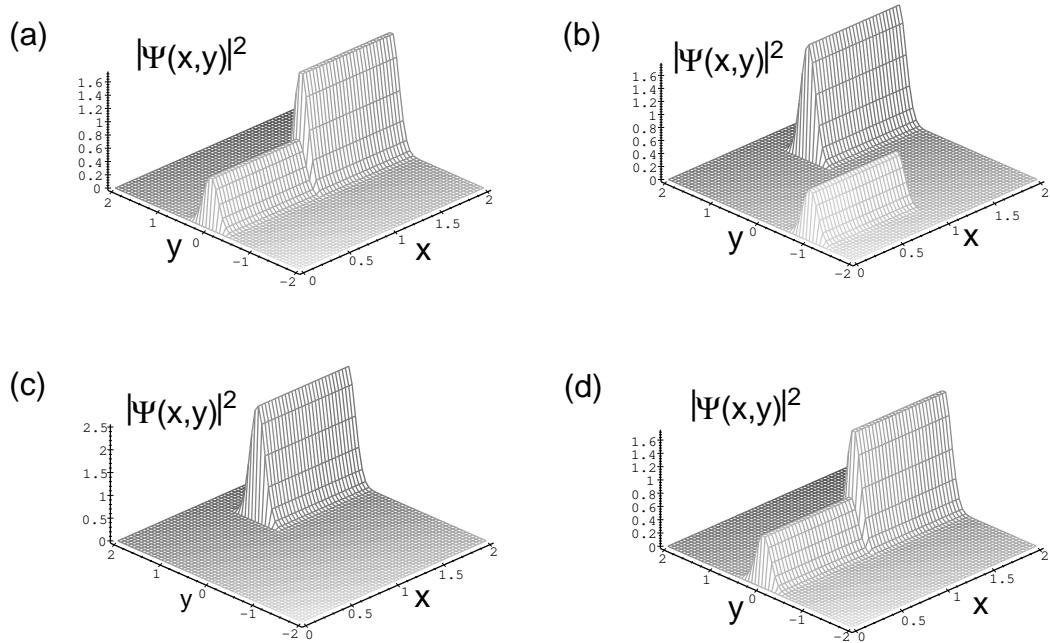}
\caption{
Measurement and repreparation:
 (a) At the initial time $t=t_0$ an arbitrary wavefunction
$\Psi(x,y,t_0)$ is to be measured by a detector with coordinate $y=0$.
(b) During the interaction between system and detector there is a separation
of two parts of the wavefunction corresponding to each interval $[0,1[,
[1,2[$.
(c) After the registration of the result of a measurement the wavepacket
without the particle (here [0,1[) does not influence the particle
dynamics anymore and can therefore be neglected.
(d) For a new measurement with the same initial state (a) the remaining wavepacket
has to flow from one to both intervals into the original form
$|\Psi(x,y,t_0)|$.  \normalsize}
\end{center}
\end{figure}

During this whole process the position $2x_n$ of the particle during the
$n-$th measurement can be formally  mapped onto a Bernoulli-shift 
$x_{n+1}= 2 x_n \; {\rm mod} \; 1$.
Thereby the rescaling with the factor $2$ is due to the flow of the
wavepacket -- and of the ergodic particle trajectory $x(t)$ respectively -- 
during the preparation of the initial state
for the next measurement. The calculation ${\rm mod}
1$ corresponds to the fact that without loss of generality only the
wavepacket containing the particle is kept after the measurement and that the 
support $M_{x_n}$ of this wavepacket can be identified with the original 
phase space. 
In other words: the series of measurements and repreparations leads to a 
B\"acker-Transformation (closely connected to a Bernoulli-shift)
of the accesible area in phase space, in which
the trajectory $x(t)$ is ergodic.  

The results obtained in the simple model presented here also hold true for 
a general position measurement with finite measurement intervals, where 
a generalized form of the Bernoullishift is to be used.  
As the Bernoulli-shift is the standard example and basic ingredient of
chaotic motion \cite{Hi94}, the motion of the hidden variable shows
deterministic chaos during any (position) measurement.

Although the position of the particle evolves deterministicaly from the
initial value $x(0)$, the intrinsic inaccuracy $\Delta x$ 
 of any measurement together with the mixing property of the dynamics 
 prevents from the complete knowledge of the system. 
Therefore in this theory the
result of a future measurement can in principle not be predicted
from the history of the system,
 although no stochastic features have been introduced in the 
theory. In this sense it
 might be that God does not play at dice, but we do not look closely 
enough to discover.

\subsection{Quantum equilibrium}

\subsubsection{Problem}

First consider an ensemble $\{ x_i \}$  of independent systems 
with the same wavefunction
$\Psi(x,t)$,
but different positions $x(t)$ of the particles (cf. Fig. 4), according to the 
distribution function $\mathcal{P}(x,t)$. The
relation $\mathcal{P}(x,t)=
\vert\Psi(x,t)\vert^2$ is called "quantum equilibrium" 
\cite{Du92a}, while the alternative
$\mathcal{P}(x,t)\neq\vert\Psi(x,t)\vert^2$ is ruled out by experimental
results (cf. Fig. 4).

\begin{figure}
\leavevmode
\begin{center}
\epsfxsize=\textwidth
\epsfbox{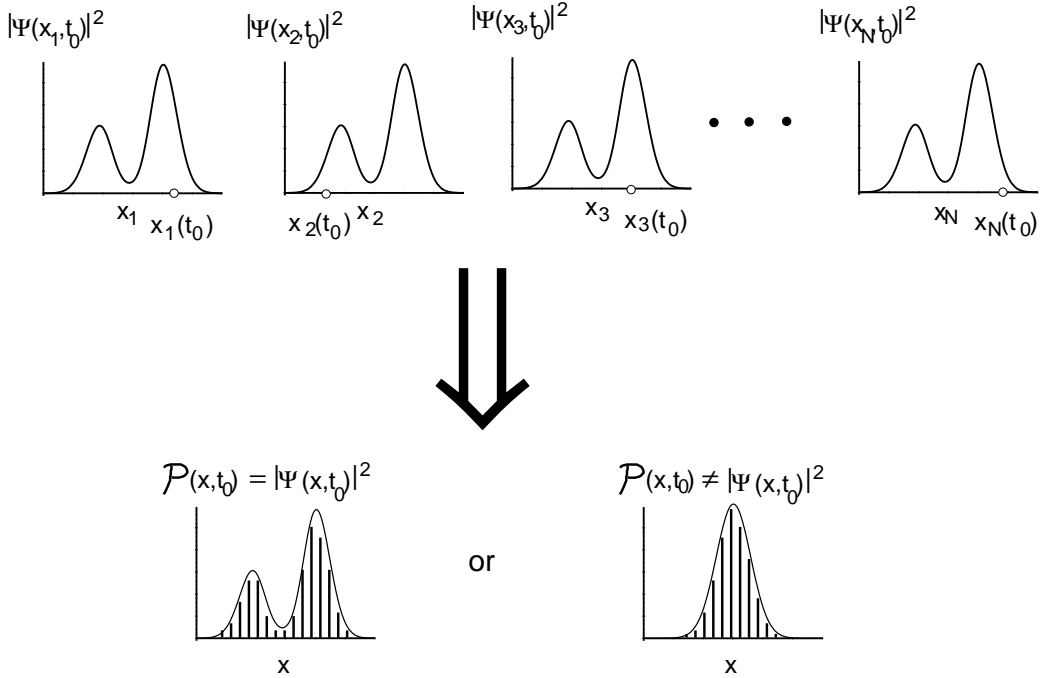}
\end{center}
\caption{
The problem of quantum equilibrium in the Bohmian quantum mechanics: 
Let an ensemble be a collection of $N$ independent systems, which have a
wavefunction $\Psi(x_i,t_0)$ with the same
$\vert\Psi(x_i,t_0)\vert^2$ for all $i\in\{1,\ldots,N\}$, the position
$x(t_0)$ of the particle being distributed according to the position density
$\mathcal{P}(x,t)$. The case
$\mathcal{P}(x,t_0)=\vert\Psi(x,t_0)\vert^2$ is called quantum equilibrium, 
which is derived in chapter \ref{derivation} from the properties of the 
measurement process.
 }
\end{figure}

Because of the analogous form of the continuity equations
\begin{eqnarray} 
0 &=& \frac{\partial }{\partial t} \mathcal{P}(x,t) +\frac \partial {\partial
x}({\mathcal{P}}(x,t) \frac{\frac{\partial}{\partial x}S(x,t)}{m})  
\label{cont1} \quad , \\
0 &=& \frac{\partial }{\partial t}\vert\Psi(x,t)\vert^2 +\frac 
\partial {\partial
x}(\vert\Psi(x,t)\vert^2 \frac{\frac{\partial}{\partial
x}S(x,t)}{m})  \label{betr} \label{cont2}
\end{eqnarray}
for
${\mathcal{P}}(x,t)$ and
$\vert\Psi(x,t)\vert^2$ it is sufficient to assume or derive $\mathcal{P}(x,t_0)=\vert\Psi(x,t_0)\vert^2$
at an arbitrary time $t_0$, in order to guarantee quantum equilibrium for
all times.

The problem is that this assumption originally made by D. Bohm \cite{Bo54,Ke53}
has  a statistical character, which is in contrast to the rest of the 
Bohmian quantum mechanics,  which is a single system theory
without any statistical inputs. 
Several propositions have been made to justify this: 
random collisions \cite{Bo53}, coarse graining \cite{Va91}, 
subquantum fluctuations \cite{Bo54,Bo93,Bo89} or properties of the total
wavefunction of the universe \cite{Du92a,Du92b,Du93}.

\subsubsection{Derivation from the measurement process \label{derivation}}

In the following this problem will be solved by considering the results of 
a sequence of measurements of a {\em single} system, which exactly 
coincides with  the real physical situation to be described.

Before each measurement
at the times $t_i$ the wavefunction is prepared again in the state  
$\Psi(x,t_0)=\Psi(x,t_i)$.
As already discussed in chapter \ref{measurment}
the registration of the detector state involves at least one position
measurement.
As it has also been proven that the particle trajectory during any
nontrivial position measurement
is chaotic (cf. chapter \ref{chaos}),
it can be concluded that the motion of the particle during
almost all nontrivial
measurements is ergodic \cite{Sch88}.

Let $P(x)$ be the distribution of the position $x_i=x(t_i)$ determined in a
sequence of measurements of a single system. Due to ergodicity of the 
trajectory $x (t) $ the distribution 
 $P(x)$, which
is obtained along one trajectory $x(t)$ at different times $t_i$, can also
be expressed by the probability distribution 
$\mathcal{P}(x,t_0)$ of an appropriate fictive ensemble of particles
 $\{x_i\}$ at
a fixed time $t_0$: 
\begin{equation} 
P(x)={\mathcal{P}}(x,t_0) \quad . \label{zwischen1}
\end{equation}
Formally this identity can be concluded from the coincidence of the time 
average 
$\langle x^k \rangle_t := \frac{1}{M} \sum_i x^k (t_i)$
and the ensemble average
$\langle x^k \rangle_e := \frac{1}{V} \int x^k dx$
in ergodic systems for all $k \in \Zset$. 
Note that here the ensemble $\{ x_i \}  $
is not introduced by an additional statistical 
assumption, but follows from the proven ergodicity of the dynamical system 
under investigation. 

It will now be demonstrated that the distribution function of any ensemble of
particles, which move according to equation (\ref{pH}), is determined
uniquely by the restriction posed by the continuity equations 
\ref{cont1} and \ref{cont2}
for ${\mathcal{P}}(x,t)$
and $\vert\Psi(x,t)\vert^2$. For technical details see 
\cite{Ge98,geigerzukunft}.

Note that
the continuity equation 
\begin{equation} \frac{\partial }{\partial t}{\mathcal{P}}(x,t) +\frac \partial {\partial
x}({\mathcal{P}}(x,t) \frac{\frac{\partial}{\partial
x}S(x,t)}{m})  = 0, \label{schl} \end{equation}
is formally identical to equation (\ref{betr}).
Defining $f(x,t)$ implicitly by
\begin{equation} {\mathcal{P}}(x,t)=f(x,t)\vert\Psi(x,t)\vert^2 \label{f}
\end{equation}
and inserting (\ref{f}) and (\ref{betr}) into (\ref{schl}) we get 
$\frac d {dt}  f(x(t),t)=0$ for particle trajectories $x(t)$,
i.e. $f(x,t)$ is constant along trajectories.
As the trajectory is ergodic, it is a dense subset of phase space and 
$f(x,t)$ is constant in the whole accessible area. Because of
the normalisation $\int {\mathcal{P}}(x,t) dx= \int \vert\Psi(x,t)\vert^2
dx=1$ it follows that $f(x,t)=1$ and
\begin{equation}
{\mathcal{P}}(x,t)=\vert\Psi(x,t)\vert^2 \quad . \label{zwischen}
\end{equation}

Finally we can conclude that the distribution $P(x)$ of the real
experimental results during a sequence of measurements in a single system is
given by  
\begin{equation} P(x)
\stackrel{{\rm equ.} \ref{zwischen1}}{=}
{\mathcal{P}}(x,t_0)
\stackrel{{\rm equ.} \ref{zwischen}}{=}
\vert\Psi(x,t_0)\vert^2= \vert\Psi(x,t_i)\vert^2 \quad ,
\end{equation}
where the last step follows from the consecutive repreparation of the system
at times $t_i$ in the inital state.

This means that the density $P(x)$ of particle positions in a sequence of
measurements is identical
to $\vert\Psi(x,t)\vert^2$ of the wavefunction of the single systems which
is prepared before each measurement.

This central result indicates that the microscopic, deterministic
dynamics of a single system 
during a sequence of measurements produces the statistical
predictions of the standard quantum mechanics without any statistical 
assumptions. 
Thereby it turns out that deterministic chaos is responsible for 
the statistics of quantum mechanics in a similar way as it provides a
microscopic 
foundation of (classical) statistical mechanics via the proof of Boltzmann's 
$H$-theorem.

\section{Conclusion}

Within the Bohmian quantum mechanics the whole measurement process including
its statistical properties can be
described for a single system as a deterministic process without the
assumption of a reduction  collapse of the wavefunction.

Deterministic chaos can be introduced to Bohmian quantum mechanics
in the same way as in classical mechanics. In particular it has been
demonstrated that the 
motion of the Bohmian particle during the measurement is intrinsically
chaotic.

From this it can be concluded that  
Bohmian quantum mechanics gives the usual statistical prediction 
of quantum mechanics without any statistical assumptions within a single
system theory.
The uncertainty in the result of a quantum mechanical measurement follows
from the interplay of the chaotic motion of the hidden variable 
$x(t)$ and the finite accuracy $\Delta x$ of any real measurement.
Also the experimentally confirmed probability postulate 
$ P (x) =\vert\Psi(x,t)\vert^2$ can be derived as a time
average of a sequence of measurements of the same system.
In the standard formulation of quantum mechanics this feature is not derived 
from a physical process, but ad hoc introduced by the reduction collapse of the
wavefunction. 
Our result shows that the appearance of deterministic chaos allows for the
derivation of all statistical properties of quantum mechanics within a causal 
formulation for a single system.

More detailed investigations of the presented results and implications for 
the classical limit of Quantum mechanics will be presented in a forthcoming 
publication \cite{geigerzukunft}.


\end{document}